\documentstyle[twocolumn,fleqn,espcrc2,psfig,afterpage]{article}
\newcommand{\AmS}{{\protect\the\textfont2
  A\kern-.1667em\lower.5ex\hbox{M}\kern-.125emS}}

\def\lsi{\raise0.3ex\hbox{$<$\kern-0.75em\raise-1.1ex\hbox{$\sim$}}}
\def\gsi{\raise0.3ex\hbox{$>$\kern-0.75em\raise-1.1ex\hbox{$\sim$}}}

 {\begin{list}{}{%
  \setlength{\leftmargin}{10pt}%
  \setlength{\itemindent}{10pt}%
  \setlength{\rightmargin}{0pt}}}
  {\end{list}}

 {\begin{list}{}{%
  \setlength{\leftmargin}{8pt}%
  \setlength{\itemindent}{8pt}%
  \setlength{\rightmargin}{0pt}}}
  {\end{list}}

\begin{document}
\onecolumn
\title{Fermions on random lattices}

\author{{{Z.~Burda$^{{\rm a  b}}$,~~J.~Jurkiewicz$^{\rm b}$,~~A.~Krzywicki 
}}\address{Laboratoire de Physique
Th\'eorique, B\^at. 210, Universit\'e 
Paris-Sud, 91405 Orsay, 
France$^{\rm 1}$ \\
$^{\rm b}$Institute of Physics, Jagellonian University, 
30-059, Krak\'ow,  Poland
 }}
       

\begin{abstract}
We put fermions and define the Dirac operator and 
spin structures on a randomly
triangulated 2d manifold.
\end{abstract}

\maketitle  
\addtocounter{footnote}{1}
\footnotetext{ Unit\'e Mixte du CNRS UMR 8627.
 Orsay rep. LPT 99-61}

\section{Introduction}

\noindent

It is known, that for more than 2d
the constructive, lattice approach towards gravity
quantization faces the problem of the apparent
instability of random manifolds. Several 
ideas are under discussion.
One is that a sensible quantum gravity theory
should be supersymmetric in the continuum limit. 
World-sheet supersymmetry is
necessarily broken on the lattice. But, perhaps, it
is sufficiently weakly broken to stabilize the manifolds.
The problem is notoriously very difficult. The first step 
in this direction is to define a spin structure on a
piecewise linear, randomly triangulated manifold. We consider
here 2d manifolds made up of equilateral triangles.
Generalizations of our method to arbitrary triangulations 
and to $d>2$ are possible, but beyond the scope of this report.
For more details see \cite{bjk}.

\section{Spin connection and spin structures}

\noindent

We consider a randomly triangulated manifold. The fermions
are assumed to live on the dual lattice. Our starting point 
is the familiar gauged Wilson action for
fermions:
\begin{equation}
S = \sum_{\langle b, a \rangle} \bar{\psi}(b) [ - \kappa 
(1 + n_{ba} \cdot \gamma) U(ba) + \mbox{$\frac{1}
{4}$} \delta_{ba}] \psi(a)\; 
\label{action}
\end{equation}
Here $b, a$ are two neighboring sites (triangles), 
$n_{ba}$ is a unit vector along the oriented link $ba$
and $\kappa$ is the hopping parameter~. $U_{ba}$
is a connection matrix which transports the spinor $\psi$
from $a$ to $b$. In the standard lattice gauge
theory $U_{ba}$ operates in the internal symmetry space.  In
quantum gravity it connects the neighbor local frames.
\par
Let us recall, that the group of general coordinate tansformations
has no spinor representation. Thus, in general relativity, in order
to define spinors one has to introduce local orthonormal frames
transforming under $SO(d)$. In our case, the manifold is piecewise
linear, and a local frame $\{e^j(a)\}$ is associated with each 
triangle $a$. The parallel transport from $a$ to $b$ along some
curve $C$ is:
\begin{equation}
e^j(b) = U(b \stackrel{\ \ C}{\leftarrow} a)^j_k \ e^k(a)
\label{vcongen}
\end{equation} 
This is to be lifted to the spinor representation
\begin{equation}
\psi^\alpha(b) =
{\cal U}(b \stackrel{\ \ C}{\leftarrow} a)^\alpha_\beta
\ \psi^{\beta}(a) \quad , \quad {\cal U} = \big[U\big]_{1/2}
\label{scongen}
\end{equation}
Here one encounters the well-known sign ambiguity: the matrix 
${\cal U}$ is defined up to sign. However, when $C$ is a 
loop, the sign of ${\cal U}(a \stackrel{\ \ C}{\leftarrow} a)$
is determined by the sign of any other loop into
which $C$ can be continuously deformed. Consequently, when 
$C$ is contractable to a point, like every loop on a sphere, 
the sign of ${\cal U}(a \stackrel{\ \ C}{\leftarrow} a)$ is 
actually determined and a unique spin structure can be defined.
For  non-contractable loops there are two
possible sign choices. Since there are two types of non-contractable
loops on a torus, four distinct spin structures can be defined on
it. This generalizes in an obvious way to orientable surfaces of 
higher genus. On the Klein bottle, however, one meets an
obstruction: the spin structure cannot be defined.
The problem of putting fermions on a random lattice boils down to
that of defining  the spin structure on it consistently.

\section{The explicit construction of the connection}

\noindent

The local frames in two neighboring triangles are connected
by $e(b)=U(ba)e(a)$,. The matrix notation is used to drop
the indices appearing in (\ref{vcongen}). To the local gauge
transformation of the frames $e(a) \rightarrow \Omega(a) e(a)$
corresponds the transformation $U(ba) \rightarrow \Omega(b)
U(ba) \Omega^{-1}(a)$. Clearly, $Tr U(C)$ is gauge invariant if $C$ is
a closed loop.
\par
Let $R(\phi) = \exp{(\epsilon \phi)}$ denote the matrix performing 
the rotation by angle $\phi$ ($\epsilon$ is the rotation generator).
It is easy to convince oneself that $U(ba)$ can be written
\begin{equation}
U(ba) = R^{-1}(\phi_{b \rightarrow a}) R(\pi) 
R(\phi_{a \rightarrow b})
\label{vcon}
\end{equation}
\begin{figure}[t]
 \centerline{\psfig{figure=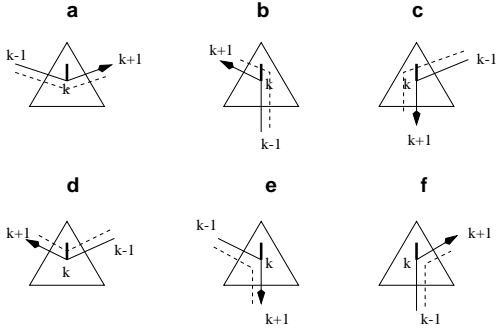,width=2.6in,angle=270}}
 \caption{Six possible ways a dual lattice
loop can go through a triangle. The line segment pointing up
from the centre of the triangle is the vertex flag
indicating the gauge choice, i.e. the direction of
the vielbein $e^1(k)$. The dashed line is the loop
slightly displaced to the right. The sign factor
in Table 1 is negative when the dashed line crosses the flag. }
\end{figure}

where $\phi_{a \rightarrow b}$ is the angle between the 1st axis
of $e(a)$ and $n_{ab}$. Likewise, $\phi_{b \rightarrow a}$ is
the axis of the 1st axis of $e(b)$ and $n_{ba}$. The angles are
oriented: they are always measured counterclockwise from $e$ to
the appropriate $n$. The parallel transporter along a closed 
loop is the product of successive rotations 
$R(\pi)R(\phi_{k \rightarrow k-1}) R^{-1}
(\phi_{k \rightarrow k+1}) = R(\pm \pi/3)$, the sign $\pm$
depending on whether the loop turns right or left. For an
elementary loop $L_q$, encircling a single vertex of the
triangulated lattice in $q$ steps, one obviously has  
\begin{equation}
\frac{1}{2} Tr U(L_q) = \frac{1}{2} Tr R(\pm \pi/3) =  
\cos(q\pi/3)
\label{vcurv}
\end{equation}
As expected the rhs of (\ref{vcurv}) equals unity when the surface is flat,
i.e. when $q=6$ (remember that the triangles are equilateral!).
For general $q$ the rhs equals the cosine of the monodromy angle
and measures the curvature.
\par
In the spinor representation we write in anology to (\ref{vcon}):
\begin{equation}
{\cal U}(ba) = g_{ba}{\cal R}^{-1}(\phi_{b \rightarrow a}) {\cal R}(\pi) 
{\cal R}(\phi_{a \rightarrow b})
\label{scon}
\end{equation}
where ${\cal R}(\phi) = \exp{(\epsilon \phi/2)}$ (in 2d the
rotation generator in both representations is the same matrix)
and $g_{ba}$ is a sign factor. Our problem, actually a topological
problem, is to fix the sign factors $g_{ba}$ all over the lattice for a
given gauge. Of course, physical quantities are gauge independent.
\par
The spinor analogue of (\ref{vcurv}) is
\begin{equation}
\frac{1}{2} Tr {\cal U}(L_q) =   
F(L_q) \cos(q\pi/6)
\label{scurv}
\end{equation}
where $F(L_q)$ is a product of $g$'s appearing in (\ref{scon})
and of additional sign factors, to be calculated in a 
moment and given in Table 1. Clearly, for $q=6$ the rhs 
 must equal unity and
therefore $F(L_6)= -1$. One can produce several argument to
show that 
\begin{equation}
F(L_q) = -1 \; \; \forall q
\label{elem}
\end{equation}
E.g. one can smear the metric singularity at the vertex within
$L_q$. For an infinitesimal loop the deficit angle is then zero
and the loop sign factor is -1. As one enlarges the loop, the
deficit angle is progressively built. The loop sign factor 
keeps its value -1, since $Tr {\cal U}(L_q)$ varies continuously.
Eq. (\ref{elem}) is an essential constraint.
\par
\begin{table}
\caption{ The
last column is ${\cal R}(\pi){\cal R}(\phi_{k \to
k-1}){\cal R}^{-1}(\phi_{k \to k+1})$ and is a
product of a specific sign factor and of an exponential, the
Kac-Ward factor.}
$$\begin{array}{crrc}
\mbox{Fig 1} & \phi_{k \to k-1} &  \phi_{k \to k+1} & \mbox{\rm
sign} \times \mbox{\rm Kac-Ward}
\\
& & & \\
a & \pi/3   & 5\pi/3 & + \exp(-  \epsilon \pi/6) \\
b & \pi     & \pi/3  & - \exp(-  \epsilon \pi/6) \\
c & 5\pi/3  & \pi    & - \exp(-  \epsilon \pi/6) \\ 
& & & \\
d & 5\pi/3 & \pi/3 & - \exp( +  \epsilon \pi/6) \\
e & \pi/3   & \pi  &  + \exp( +  \epsilon \pi/6) \\
f & \pi  & 5\pi/3  &  + \exp( +  \epsilon \pi/6)  
\end{array}$$
\end{table}
The calculation of the rhs of (\ref{scurv}) is similar to
that leading to (\ref{vcurv}). Again the 
loop turns by $\pm \pi/3$ modulo $2\pi$, but now the $2\pi$
is not innocent, since in the spinor representation it is
the half-angle that matters. This can possibly yield an extra
sign factor, which depends on how the dual loop goes through the
successive triangles, say $k-1, k$ and $k+1$. It is covenient
to choose the gauge where $e^1(k)$ points from the center of
the triangle $k$ towards one of its vertices. The six possible
cases are shown in Fig. 1. The relevant result is given in 
Table 1. The trace of the product of the Kac-Ward factors
$\exp( \pm \epsilon \pi/6)$ yields $\cos{(q\pi/6)}$. The
loop factor $F(L_q)$ is a product of $g$'s and of the sign
factors to be read from Table 1. The global gauge choice can be
represented graphically, by attaching flags to dual vertices and
links. A vertex flag is just the direction of $e^1$ (one can turn
it to the right or to the left, provided one does not cross
any dual link). A link flag is put on the right of the link $ba$
if $g_{ba} = -1$. It follows, that the topological constraint
(\ref{elem}) is satisfied if the number of flags inside every
elementary loop is odd.

\section{Loop signs and topology}

\noindent

The flags are a useful tool, which helps proving the following
two theorems:
\par
- {\bf T1 }: For an arbitrary orientable triangulation one can
choose the gauge, i.e. the orientations
 of the local frames and the link
sign factors, so that $F(L) = -1$ for every elementary
loop $L$.
\par
The idea of the proof consists in checking first the validity
of T1 for a minimal sphere. Then one verifies that an ergodic
triangulation building algorithm is compatible with the 
theorem. Finally, by gluing spheres in an appropriate
manner one extends the result to a sphere with handles.
\par
 - {\bf T2 } : One has $F(C) = -1$ for every contractable loop.
\par
This can be checked by gluing elementary loops.   

\section{Wilson fermions on a randomly triangulated
manifold}

\noindent

One has 
\begin{equation}
\gamma_{ba} \equiv n_{ba} \cdot \gamma =
{\cal R}^{-1}(\phi_{b \rightarrow a}) \gamma^1
{\cal R}(\phi_{b \rightarrow a})
\end{equation}
According to (\ref{action}) the mass independent part of the
Dirac operator is
\begin{equation}
D(ba) = \frac{1}{2}(1+\gamma_{ba}) {\cal U}_{ba}
\end{equation}
Hence, finally
\begin{equation}
D(ba) = g_{ba} {\cal R}^{-1}(\phi_{b \rightarrow a})
\frac{1}{2}(1+\gamma^1) {\cal R}(\phi_{a \rightarrow b})
\label{dirac}
\end{equation}
The Dirac operator is constructed once
all the flags are put. This determines the angles and the $g$'s.
With our choice of gauge the angles $\phi$ take the values
$\pi/3, \pi, 5\pi/3$. Thus, there are nine possibilities for
the matrix on the rhs of (\ref{dirac}). These nine matrices
can be calculated beforehand.
\par
Using (\ref{dirac}) one readily calculates the 
Majorana fermion loop expectation value
\begin{equation}
\langle \bar{\psi}(1) D(12)  ... 
D(n1) \psi(1) \rangle = - F(C) (\sqrt{3}/2)^n
\end{equation}
When the loop is contractable
 $F(C)= -1$ and the rhs is just $(\sqrt{3}/2)^n$. Since
there is a one-to-one correspondence between contractable
loops and Ising spin domain boundaries, this result can be
used \cite{bm} to demonstrate the equivalence of Ising spins and
Majorana fermions on a sphere. The result can be extended to
arbitrary orientable surfaces, provided in the fermion
theory one sums over all possible spin structures. Only then the
contribution of unpaired non-contractable fermion loops cancels. The
similarity between this prescription and the GSO projection has
been noted by Polyakov \cite{pol}.

\end{document}